\documentclass[journal=jacsat,manuscript=article]{achemso}

\usepackage[version=3]{mhchem} 

\author{Mbaye Diouf}
\affiliation[Brown University]
{School of Engineering, Brown University, Providence, RI, 02912, USA}
\altaffiliation{These authors contributed equally}
\author{Joshua A. Burrow}
\affiliation[Brown University]
{School of Engineering, Brown University, Providence, RI, 02912, USA}
\altaffiliation{These authors contributed equally}
\author{Krishangi Krishna}
\affiliation[Brown University]
{School of Engineering, Brown University, Providence, RI, 02912, USA}
\author{Rachel Odessey}
\affiliation[Brown University]
{School of Engineering, Brown University, Providence, RI, 02912, USA}
\author{Ayman F. Abouraddy}
\affiliation[CREOL]
{CREOL The College of Optics \& Photonics, University of Central Florida, Orlando, FL 32816, USA}
\author{Kimani C. Toussaint Jr.}
\affiliation[Brown University]
{School of Engineering, Brown University, Providence, RI, 02912, USA}
\email{kimani_toussaint@brown.edu}

\title[An \textsf{achemso} demo]
{Excitation of Surface Plasmon Polaritons by Diffraction-free and Vector Beams}

\abbreviations{IR,NMR,UV}

\begin{document} 










\begin{abstract}
Surface plasmon polaritons (SPPs) are traditionally excited by plane waves within the Rayleigh range of a focused transverse magnetic (TM) Gaussian beam. Here, we investigate and confirm the coupling between SPPs and two-dimensional Gaussian and Bessel-Gauss wave packets, as well as one-dimensional light sheets and space-time wave packets. We encode the incoming wavefronts with spatially varying states of polarization then couple the respective TM components of radial and azimuthal vector beam profiles to confirm SPP polarization-correlation and spatial-mode selectivity. Our results do not require material optimization or multi-dimensional confinement via periodically corrugated metal surfaces to achieve coupling at greater extents. Hereby, outlining a pivotal, yet commonly overlooked, path towards the development of long-range biosensors and all-optical integrated plasmonic circuits.
\end{abstract}

\section*{Introduction}
Surface plasmons polaritons (SPPs) continue to be a topic for scientific exploration and technological development since their observation more than one hundred years ago by Wood \cite{Wood_1902}. Under the appropriate conditions imposed on polarization, energy, and momentum, an optical field can couple to a surface plasmon confined within a subwavelength from the metal-dielectric interface. \cite{ZAYATS2005131,Pitarke_2006,Barnes2003} This phenomenon has been particularly exploited for sensor applications because of the high sensitivity of the resonance condition to the surrounding medium \cite{HOMOLA1999_SPR_sensors_review,Homola08_SPP_Sensor_Chem_Bio,LIEDBERG1983299}. The first criterion for exciting SPPs is that the input optical field be transverse-magnetic (TM) polarized; no surface modes exist for transverse-electric (TE) polarization. The second criterion, determined by the SPP dispersion relationship at a metal-dielectric interface, is given by
\begin{equation}
    k_{x} =  k_{SPP} =\bigg( \dfrac{\omega}{c} \bigg)\sqrt{\frac{\varepsilon_m \epsilon_d}{\varepsilon_m + \varepsilon_d}}
\end{equation}

\noindent where the complex valued dielectric function of the metal $\varepsilon_{m}$ and lossless dielectric $\varepsilon_{d}$ are expressed as functions of the angular frequency of light $\omega$ as $\varepsilon_{m}(\omega) = \varepsilon'_{m}(\omega) + i\varepsilon''_{m}(\omega)$ and $\varepsilon_{d}(\omega) = \varepsilon'_{d}(\omega)$ and $c$ is the speed of light in vacuum.\cite{Piliarik:09} SPPs exist in the plasmonic regime of a material, and so a further condition is that $\varepsilon'_{m}(\omega)< \varepsilon'_{d}(\omega)$; where the noble metals Ag and Au are excellent conductors exhibiting a large negative real dielectric value.

Often, the Drude model for metals serves as a suitable approximation at longer wavelengths away from the plasma frequency; however, frequency dependent material properties should be characterized with ellipsometry techniques after a sample has been fabricated. It has been observed that SPPs exhibit longer propagation lengths for low-loss plasmonic systems. \cite{Baburin:18,Berini:19} An incoming TM polarized wavefront that meets the momentum criterion established by the SPP dispersion relationship will be excited on the metallic surface at each transverse position along the wavefront. Importantly, while the majority of reported experiments and models of SPPs assume excitation by either a plane wave or Gaussian intensity distribution, other electromagnetic waves that satisfy the wave equation have been explored. \cite{Man2015, Zhan06}. 

The Christodoulides group previously showed theoretically that SPP excitation by an Airy beam, which exhibits one-dimensional (1D) diffraction-free propagation, can result in Airy-SPPs that resist spatial spreading in one transverse direction while freely propagating along the interface, albeit along a curved trajectory.\cite{Salandrino:10} A key insight here is that the SPP excitation process requires phase matching, and therefore, the resulting surface wave should retain the attributes of the excitation field. Recently, it has been shown theoretically that SPPs excited by propagation-invariant space-time (ST) wave packets propagate at the metal-dielectric interface both dispersion and diffraction free.\cite{STSPP_numerical} ST wave packets owe their spatiotemporal propagation invariance, including self-healing, to engineered classical entanglement; specifically, the nonseparable correlation between the spatial and temporal frequencies.\cite{Kondakci:16,Kondakci2017,Kondakci:18,Kondakci2019,Bhaduri:18,Bhaduri19,Diouf21_STvectorbeams,Yessenov19_maxgroupdelaySTbeams} So-called striped ST-SPPs have been recently observed experimentally at a metal-dielectric interface by illuminating a nano-slit for SPP coupling.\cite{Ichiji22_STSPParxiv}

It is clear from these examples that the coherent nature of the SPP excitation process offers an opportunity to tailor the properties of the generated SPPs. In this work, we continue this line of investigation by exploring the influence of the input intensity distribution on the resulting SPPs, for various types of wave packets. Specifically, using a simple metal-dielectric interface comprising of thin Ag film on glass in a Kretschmann configuration, we experimentally explore the SPP response for Bessel-Gauss, light-sheet (LS), and space-time light-sheet (ST-LS) illumination and compare the results to conventional two-dimensional (2D) Gaussian illumination. Using leakage radiation microscopy,\cite{DREZET08_Leakageradiation} where the scattered SPP response is measured in the far-field, we find that the generated SPPs follow the spatial distribution of the illuminating field in all cases. As a result, we observe that the SPPs can extend the full extent of the input illumination field, which is several centimeters for the ST-LS.  In addition, we confirm that illumination by radially or azimuthally polarized vector beams produces similar effects, modified by the strict correlation between the spatial and polarization degrees-of-freedom and the requirement for TM illumination. The approach to shaping the SPP response presented here is punctuated by the simplicity of our far-field imaging setup, and the fact that our plasmonic sample requires no lithographic patterning protocols to achieve spatial shaping of the plasmonic response. 
 \begin{figure*}[ht!]
    \centering
    \includegraphics[width = 1.00\linewidth]{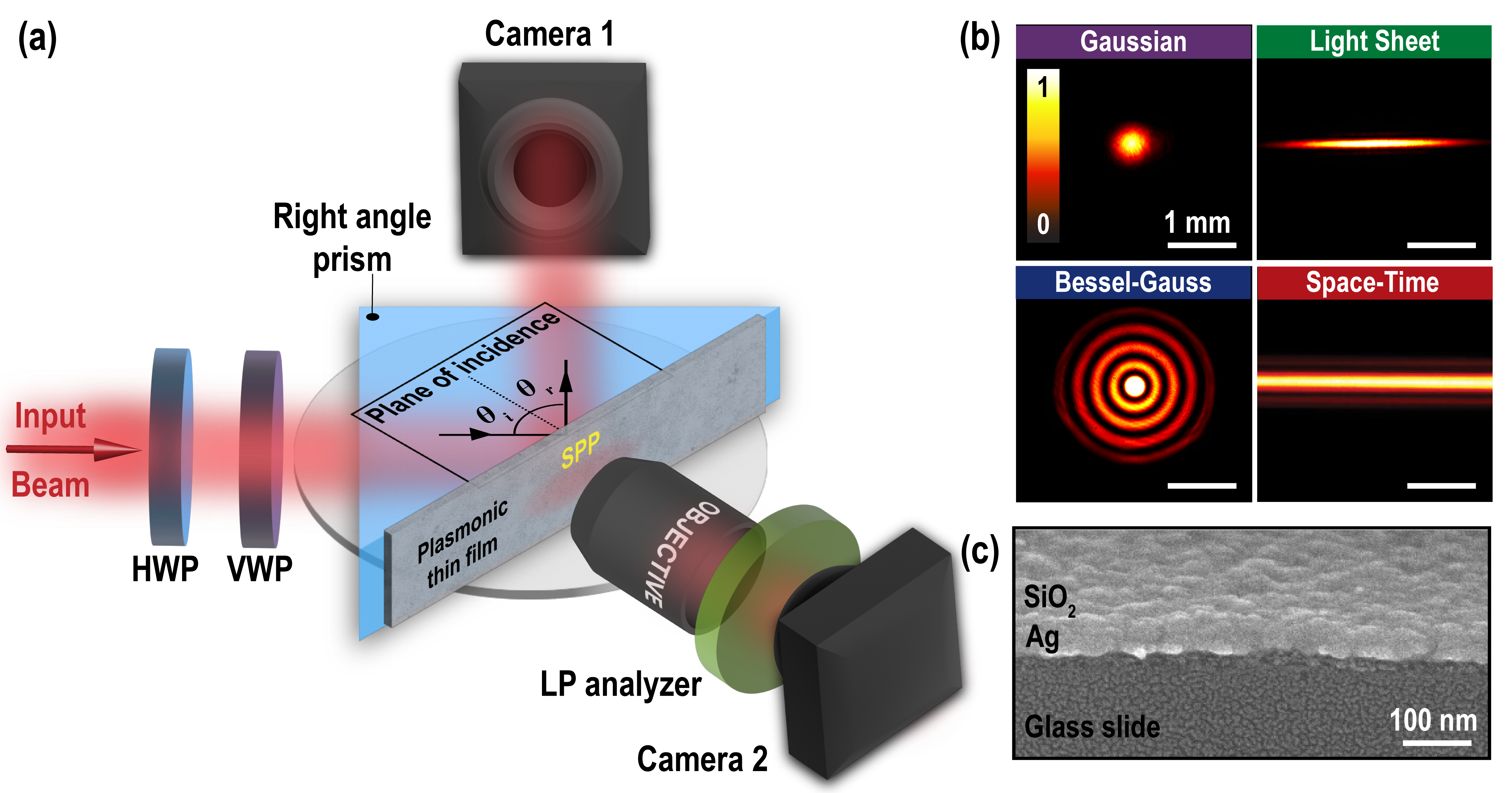}
    \caption{(a) Schematic of the experimental setup for selective SPP excitations into a homogeneous thin film configuration mounted on a right-hand prism. The total internally reflected light is imaged on Camera 1 and the scattered SPPs on the metal surface are imaged by a 5X/0.12 NA objective on Camera 2. (b) Input intensity profiles of the Gaussian, Bessel-Gauss, light sheet, and Space-Time wave packets. (c) Tilted cross-sectional SEM of the planar Ag sample.}
    \label{fig:Fig1_ExpSetup}
\end{figure*}
\section*{Experimental Section}
\textbf{Experimental setup.} Figure \ref{fig:Fig1_ExpSetup}(a) shows a schematic of the experimental setup. SPP excitation is carried out using a right-angle-prism (Thorlabs, PS913) mounted on a rotational stage (Thorlabs, XRR1) in a Kretschmann configuration, where a femtosecond-pulsed laser source, with a Gaussian intensity profile, and operating at a central wavelength of 798.4 nm is used. The source is converted to exhibit a Bessel-Gauss intensity profile by an Axicon lens (Thorlabs, AX2520). The 1D LS wave packet is synthesized through the combination of a variable mechanical slit (Thorlabs, VA100/M) and a pair of 10-mm focal length 1D cylindrical lenses for collimation. Details outlining the generation of the ST wave packets are provided in the Materials and Methods section, and more information is provided in our previous work. \cite{Diouf21_STvectorbeams} We illuminate our samples with an average power of 100 $\mu$W for all wave-packets.  The half-wave plate (HWP) in the setup is used to control the incident polarization state (TE or TM) prior to illuminating the prism. The vortex wave plate (VWP) is implemented specifically for the generation of radial and azimuthal vector beams. Note that a 30-mm focal length spherical lens (not shown) is used to focus the Gaussian and Bessel wave packets such that a beam waist of 200 $\mu$m is equal to the width of the 1D LS and the main lobe of the LS and ST wave packets. We use an sCMOS (Hamamatsu, ORCA Flash C15440-20UP) denoted by Camera 1, to image the far-field intensity profiles. We capture the reflected angular spectrum with a 1.5 x 1.5 cm$^{2}$ aperture power meter (Newport, PD 300-MS). Next, the scattered light from the SPP waves is imaged by a 5X/0.12 NA objective lens to a CMOS detector (DMK33UX178, ImagingSource, denoted by Camera 2 in Fig. \ref{fig:Fig1_ExpSetup}).  A flip mirror is placed after the imaging objective to direct the scattered light of the SPPs into a fiber-coupled spectrometer not shown (Ocean Insight, HR4000) to measure the spectral content of the electromagnetic radiation. 

\noindent \textbf{Sample preparation.} Electron beam evaporation is used to deposit a 30-nm-thick layer of silver (Ag) onto a cover glass slide that was previously coated with a 3-nm-thick titanium (Ti) wetting layer. A subsequent 20-nm-thick capping layer of SiO$_2$ is deposited to prevent oxidation of the surface. Optical thin-film thicknesses are confirmed using profilometry and cross-sectional scanning electron microscopy (SEM). Figure \ref{fig:Fig1_ExpSetup}(c) depicts a cross sectional scanning electron micrograph (SEM) of a fabricated thin film with a slight tilt to view the surface roughness of the polycrystalline metal. The sample is subsequently mounted to the right-handed prism with index-matching oil, which is adhered on a rotation mount to sweep the angle of incidence (AOI). Samples are illuminated at an angle of incidence $\theta_i$ larger than the critical angle for total internal reflection. This allows for the incoming light to have sufficient momentum for coupling to the surface plasmons. 

\section*{Results and Discussion}
 \begin{figure*}[ht!]
    \centering
    \includegraphics[width = 1.00\linewidth]{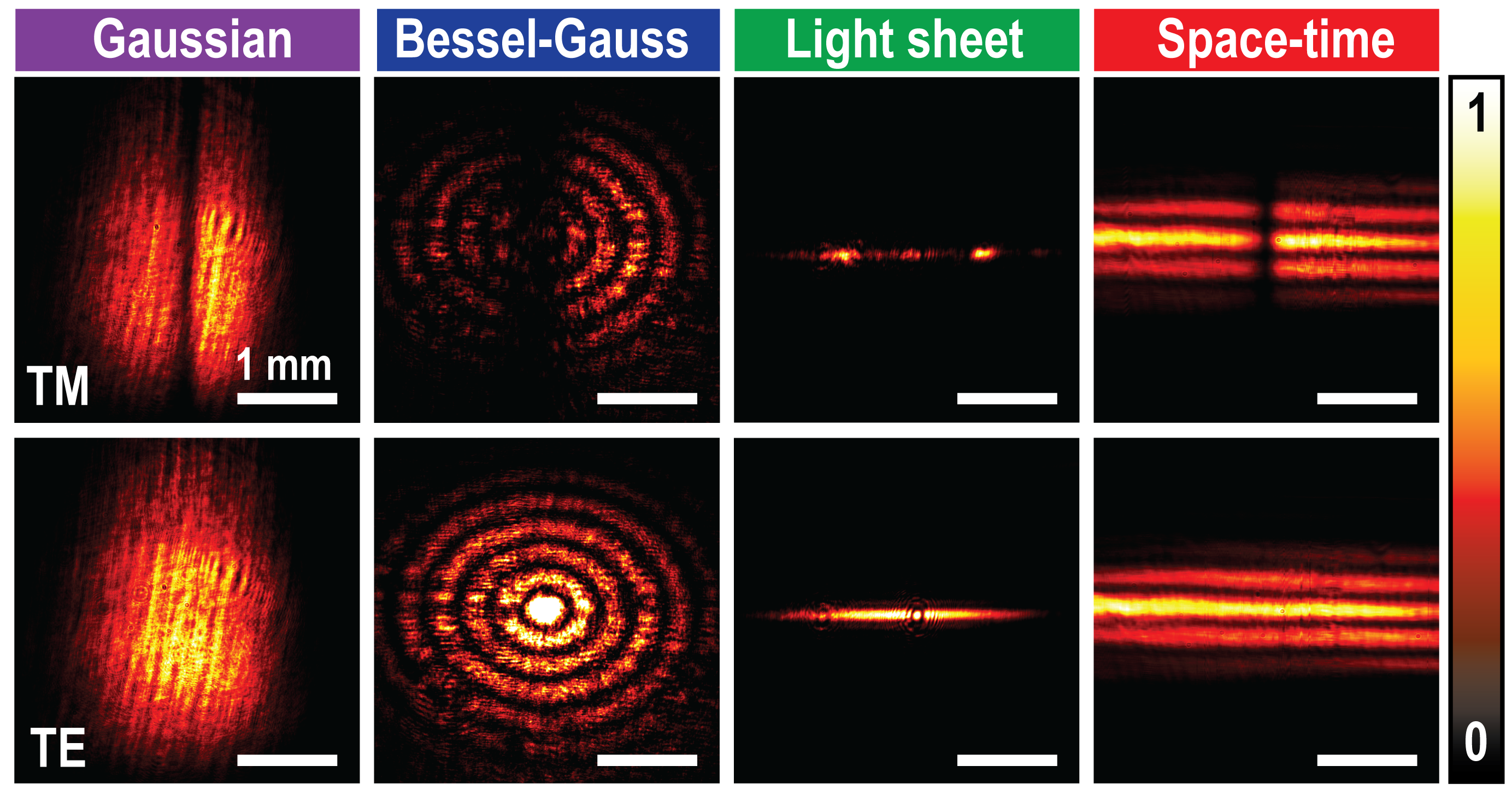}
    \caption{Reflected intensity profiles of various input intensity distributions including Gaussian, Bessel-Gauss, light sheet, and ST light sheet, for TM (top) and TE (bottom) excitations. }
    \label{fig:Fig2_v2}
\end{figure*}
\noindent \textbf{Wavefronts with spatially uniform polarization and varying intensity distributions.} To confirm the presence of the SPP, we first image the reflected wavefront at an AOI of 43$^{\circ}$ where the null is centered in the intensity profile as shown in Fig. \ref{fig:Fig2_v2}. The reflected intensity profile reveals a notable polarization dependent null at the center of the wavefront for only TM excitation (top row of Fig. \ref{fig:Fig2_v2}). This is often interpreted as coupling of the input radiation to the plasmonic system. We confirm this null is not present for TE excitation (bottom row of Fig. \ref{fig:Fig2_v2}). As the AOI deviates from the 43$^{\circ}$ resonant angle, the null horizontally translates along the wavefront. We quantitatively analyze the angular dependence on the SPP by plotting the reflected angular spectra in  supplemental document Fig. S1. The resulting spectra exhibit asymmetric line shape profiles centered at 43$^{\circ}$ with varying angular spectral widths.  For the Gaussian and the Bessel-Gauss wave packets, we observe a minimum reflectance of 18.9\% nd 31.5\%, respectively. Unlike the Gaussian and Bessel-Gauss beam distributions, the light sheet and the ST light sheet have comparable modulation depths (MD) around 42\% (see supplemental document Fig. S1). The difference in MD can be attributed to the unique spatial extent of each wavefront where the 1D wavefronts are spread over larger areas. 

Figure 3(a) shows the far-field radiation patterns at the surface of the Ag surface upon being subjected to each TM polarized wave packet. It has previously been shown that, under suitable conditions, this type of leakage radiation microscopy could be used to obtain reasonable proxies for near-field SPP propagation with the caveat that a direct 1:1 mapping cannot be obtained with this approach because of information lost in the far-field.\cite{Wiecha19_nearfieldSPP} However, we find this method attractive for its simplicity and suitability as a relative comparison among the four waves using the same sample setup. Overall, we observe an SPP-scattered intensity pattern that follows the incident wavefront. For the 1D LS and ST-LS, we only display one frame of the scattered SPPs. However, the spatial extent of the SPP follows the extent of the incident illumination field, which was apodized to 1.5 cm for our experiments. 


 \begin{figure*}[ht!]
    \centering
    \includegraphics[width = 1.00\linewidth]{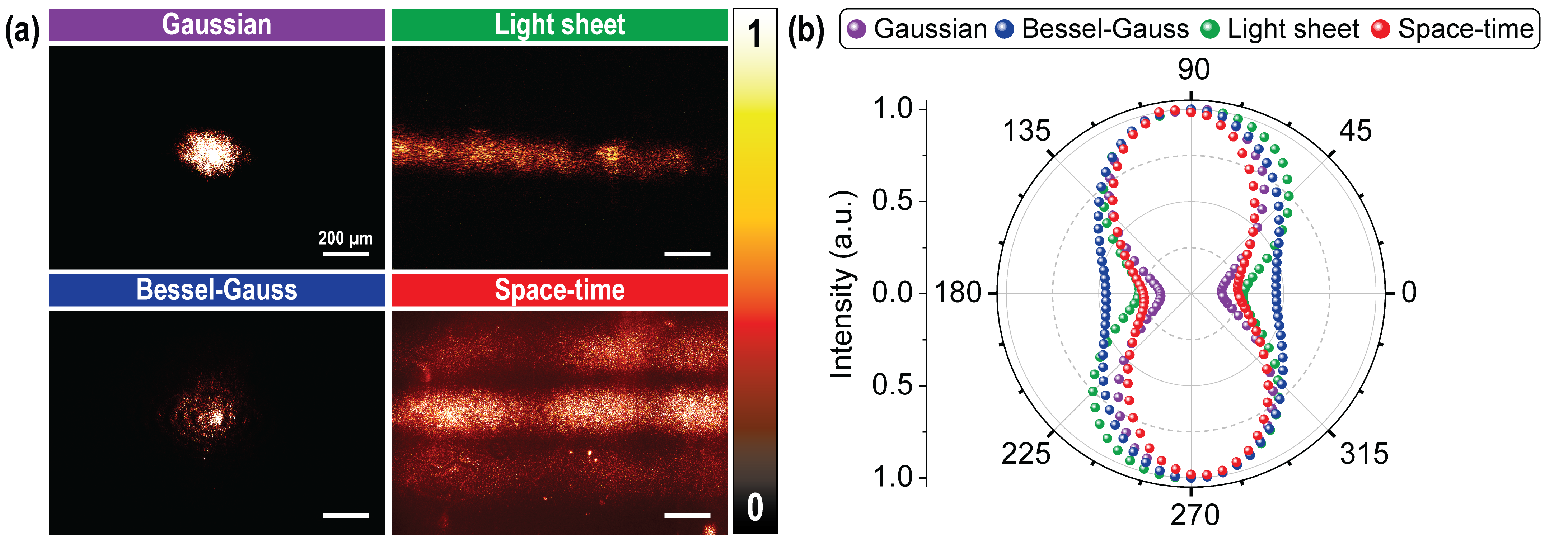}
    \caption{(a) Experimentally obtained SPP response under Gaussian, Bessel-Gauss, light sheet, and ST illumination on Ag. (b) Measured radar plot as a function of the polarization-analyzer angle derived from the resulting scattered SPP for different illumation beams.}
    \label{fig:Fig3_SPP}
\end{figure*}

Next, to examine and confirm the polarization state of the scattered SPP, we rotate the polar angle of the LP analyzer in 5$^{\circ}$ increments. For each wave packet, the scattered radiation follows a comparable dipole radiation pattern. Figure 3(b) plots the measured mean pixel intensity as a function of polar angle $\phi$ for the Gaussian (purple), Bessel-Gauss (blue), LS (green), and ST (red) scattered SPPs. We define a degree of linear polarization (DoLP) by the Stokes parameters DoLP = $\dfrac{S_1}{S_0}$, where $S_{1} = I_{90^{\circ}} - I_{0^{\circ}}$, $S_{0} = I_{90^{\circ}} + I_{0^{\circ}}$, here $I_{0^{\circ}}$ and $I_{90^{\circ}}$ are the integrated intensity values of images captured with the LP parallel and perpendicular to the scattered SPPs. We find the Gaussian, Bessel-Gauss, LS, and ST SPP wave packets to exhibit DoLPs of 0.72, 0.37, 0.56, and 0.60, respectively. Additionally, the scattered light is directed into a fiber-coupled spectrometer to measure the spectral content of the electromagnetic radiation. By applying a Gaussian fit to the measured spectra for both wave packets shown in supplemental document Fig. S2, the spectral FWHM could be extrapolated. The measured FWHM of the (a) G-SPP and the (b) ST-SPP are $\sim$8.5 nm and $\sim$3 nm centered around 800 and 798 nm, respectively.\\

\begin{figure*}[ht!]
    \centering
    \includegraphics[width = 1.00\linewidth]{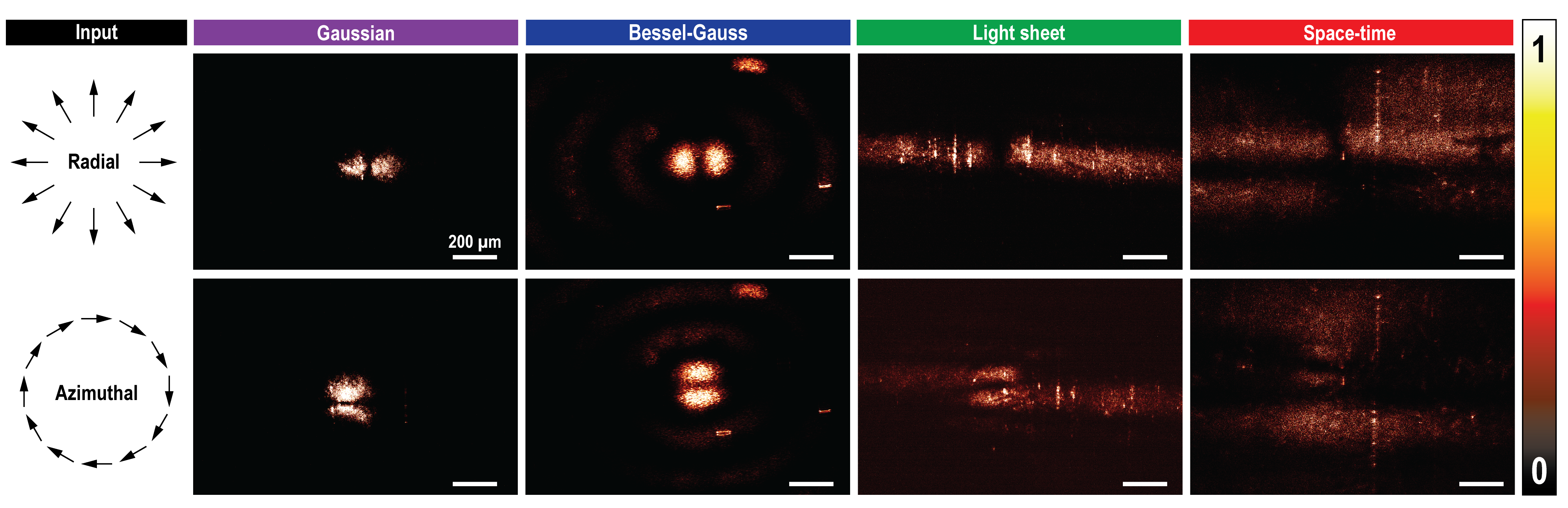}
    \caption{Detected SPP response for radially polarized (first row) and azimuthally polarized (second row) illumination using the vector-beam version of the Gaussian beam, Bessel Gauss, light sheet, and ST light sheet.}
    \label{fig:Fig4_SPPscat_Schematic}
\end{figure*}

\noindent \textbf{Wavefronts with spatially uniform polarization and varying intensity distributions.} In the above, we have excited the SPPs with various intensity distributions, each possessing a uniform polarized field. Next, we excite SPPs using vector-beam illumination, which exhibit spatially varying polarization distributions. Standard cylindrical vector beam wave-packets can be expressed as the superposition of orthogonal Hermite-Gaussian HG$_{01}$ and HG$_{10}$ modes 
\begin{align}
    \vec{E}_{r} = \text{HG}_{10}\hat{x} + \text{HG}_{01}\hat{y}\\
    \vec{E}_{\phi} = \text{HG}_{01}\hat{x} + \text{HG}_{10}\hat{y}
\end{align}
where $\vec{\text{E}}_{r}$ and $\vec{\text{E}}_{\phi}$ denote radial and azimuthal polarization states, respectively\cite{Zhan09_cylindricalVBs,Toussaint05}. Figure \ref{fig:Fig4_SPPscat_Schematic}(a) shows the experimentally observed SPP excitation intensity distributions for both radially and azimuthally polarized vector Gaussian, vector Bessel-Gauss, vector light sheet and vector ST SPPs, as illustrated in first row and second rows, respectively. The input states of polarization are illustrated in the first column. 

As expected, for all illumination fields, SPP excitation only occurs for the TM component of the illumination field, which for vector beams are correlated, or ( ‘classically entangled’ (28), with specific spatial modes. Thus, we demonstrate that radially polarized illumination leads to SPP excitation for the HG$_{10}$ mode, while azimuthally polarized illumination results in SPP excitation for the HG$_{01}$ mode. Note that our illumination approach is without the use of tight focusing, contrary to typical procedures what is typically done. Indeed, it is known that focusing radially polarized light with a high-NA objective produces a strong longitudinal field, which satisfies the polarization condition for SPP excitation \cite{Zhan06, Man2015}.

\begin{figure}[ht!]
    \centering
    \includegraphics[width = 0.5\linewidth]{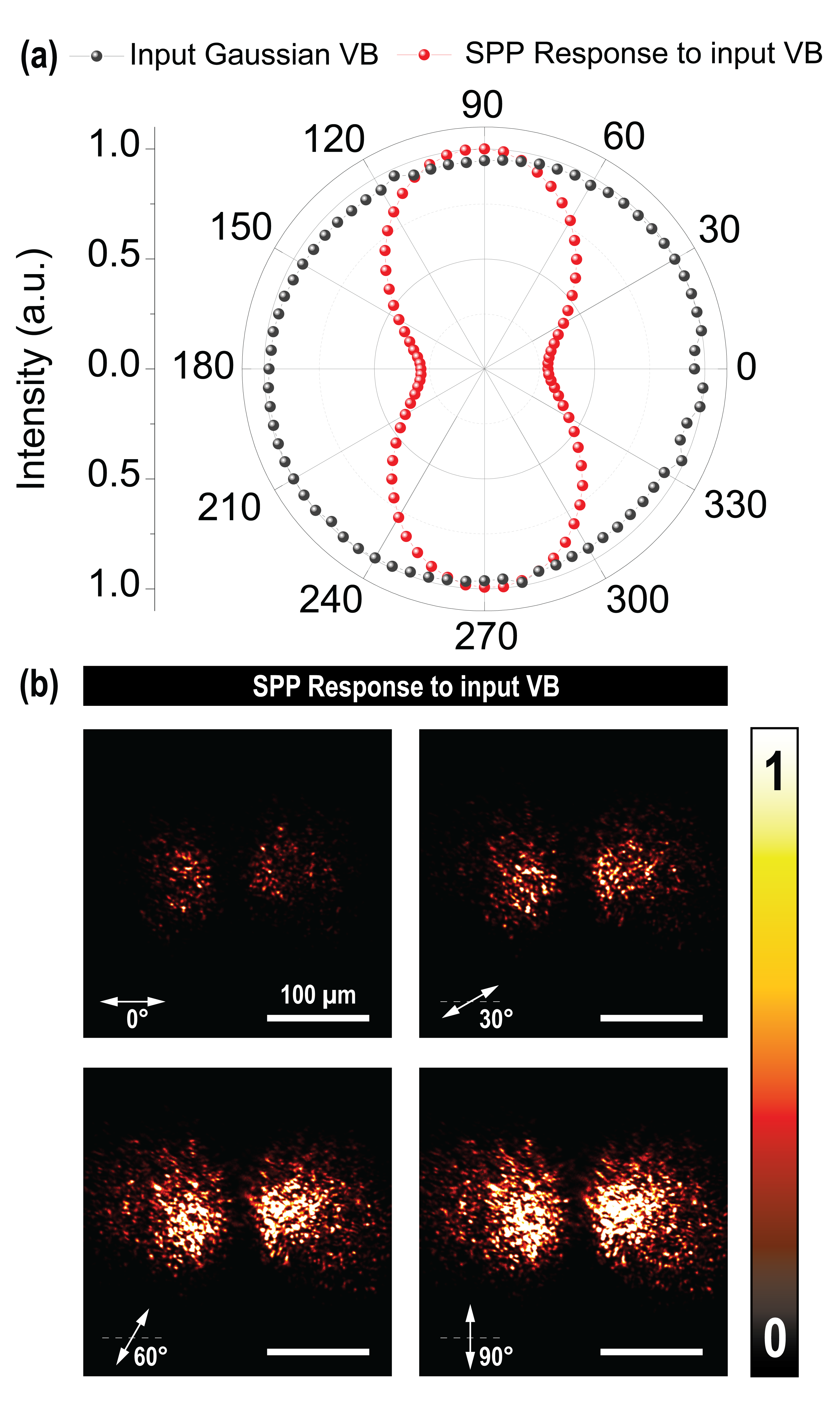}
    \caption{(a) Radar plot comparing the input radially polarized Gaussian vector wave-packet and its corresponding SPP response. (b) The measured SPP response to the input radially polarized beam projected along the horizontal, 30$^{\circ}$, 60$^{\circ}$, and 90$^{\circ}$ directions of the polarization analyzer. }
    \label{fig:Fig5_VBanalyzer}
\end{figure}

Figure \ref{fig:Fig5_VBanalyzer}(a) depicts the transmitted intensity as a function of rotation angle of the LP analyzer measured in 5$^{\circ}$ increments for the radially polarized input Gaussian beam (black) and the resulting SPP response (red). As expected, the input vector beam shows azimuthal symmetry with rotation of the polarization analyzer. Conversely, the resulting SPP response exhibits the familiar HG$_{10}$  mode that does not rotate with the polarization analyzer and possesses a DoLP = 0.55. The intensity distribution of the SPP response after passing through different orientations of the analyzer are depicted in Fig. \ref{fig:Fig5_VBanalyzer}(b). The orientations of the analyzer are depicted by the white arrows.

\section*{Conclusion}
In conclusion, we have employed a series of wave-packets with unique transverse intensity profiles, namely a Gaussian, Bessel-Gauss, LS, and ST, into a plasmonic system capable of supporting SPP modes. The thin film configuration enables us to image the leakage radiation with a far-field optical setup. The far-field scattering from the coupled SPP electrons is visualized and investigated in terms of the spatial extent and scattered polarization. Additionally, we use a series of Hermit-Gaussian modes for each wave-packet to selectively couple to SPPs in regions where the transverse magnetic polarization criterion is met. 

\section*{Methods and Materials}

\subsection*{Space Time wave packet generation}
A femtosecond laser (InSight X3, Spectra-Physics) is used to generate a Gaussian beam with 150 fs pulses at a repetition rate of 80 MHz and an average power of 1 mW. The excitation wavelength for this study is centered at 798.4 nm, and the laser is spectrally adjustable between 680 and 1300 nm. The details of the experimental generation for the synthesis and characterization of the ST light sheets are depicted in our recent paper \cite{Diouf21_STvectorbeams}. For the ST-SPPs shown in Fig. 3B, we employ the subluminal regime ($v_g = c \tan \theta$) with $\theta = 44.96^\circ$ corresponds to spectral tilt angles in the range of $0^\circ < \theta < 45^\circ$. The superluminal of ST-SPPs corresponds to spectral tilt angles of $45^\circ < \theta < 180^\circ$, with a positive $v_g$. We measured the temporal bandwidth of the ST light sheet $\Delta\lambda$ = 2 nm (extracted from the spectrum of the pulsed laser see Fig. S3 in the supplemental material), the spatial bandwidth $\Delta k_x = 18$ rad/mm and the spectral uncertainty $\delta\lambda$= 50 pm with the selected spectral tilt angle $\theta = 44.96^\circ$. These parameters are the fundamental factors for preserving the ST light sheet and avoiding spatial spreading. Indeed, a broader spectrum is needed to protect a beam with larger spatial bandwidth. Sub-wavelength ST-SPPs can propagate invariantly for significant distances, according to the theory (24). Figure 3A and 3B shows the differences between a standard G-SPP and an ST-SPP, both with an initial transverse width of 8 $\mu$m (FWHM).

\section*{Acknowledgments}
We acknowledge the Center for Nanoscale Systems (CNS) at Harvard University for support with sample fabrication. We thank Dr. Nick Fang for helpful discussions. J.A.B. would like support from the Hibbitt Postdoctoral Fellowship award.  We thank also Jieliyue (Jillian) Sun for assisting with experimentation and data analysis.

\bibliography{scibib}

\providecommand{\latin}[1]{#1}
\makeatletter
\providecommand{\doi}
  {\begingroup\let\do\@makeother\dospecials
  \catcode`\{=1 \catcode`\}=2 \doi@aux}
\providecommand{\doi@aux}[1]{\endgroup\texttt{#1}}
\makeatother
\providecommand*\mcitethebibliography{\thebibliography}
\csname @ifundefined\endcsname{endmcitethebibliography}
  {\let\endmcitethebibliography\endthebibliography}{}
\begin{mcitethebibliography}{28}
\providecommand*\natexlab[1]{#1}
\providecommand*\mciteSetBstSublistMode[1]{}
\providecommand*\mciteSetBstMaxWidthForm[2]{}
\providecommand*\mciteBstWouldAddEndPuncttrue
  {\def\EndOfBibitem{\unskip.}}
\providecommand*\mciteBstWouldAddEndPunctfalse
  {\let\EndOfBibitem\relax}
\providecommand*\mciteSetBstMidEndSepPunct[3]{}
\providecommand*\mciteSetBstSublistLabelBeginEnd[3]{}
\providecommand*\EndOfBibitem{}
\mciteSetBstSublistMode{f}
\mciteSetBstMaxWidthForm{subitem}{(\alph{mcitesubitemcount})}
\mciteSetBstSublistLabelBeginEnd
  {\mcitemaxwidthsubitemform\space}
  {\relax}
  {\relax}

\bibitem[Wood(1902)]{Wood_1902}
Wood,~R.~W. On a Remarkable Case of Uneven Distribution of Light in a
  Diffraction Grating Spectrum. \emph{Proceedings of the Physical Society of
  London} \textbf{1902}, \emph{18}, 269--275\relax
\mciteBstWouldAddEndPuncttrue
\mciteSetBstMidEndSepPunct{\mcitedefaultmidpunct}
{\mcitedefaultendpunct}{\mcitedefaultseppunct}\relax
\EndOfBibitem
\bibitem[Zayats \latin{et~al.}(2005)Zayats, Smolyaninov, and
  Maradudin]{ZAYATS2005131}
Zayats,~A.~V.; Smolyaninov,~I.~I.; Maradudin,~A.~A. Nano-optics of surface
  plasmon polaritons. \emph{Physics Reports} \textbf{2005}, \emph{408},
  131--314\relax
\mciteBstWouldAddEndPuncttrue
\mciteSetBstMidEndSepPunct{\mcitedefaultmidpunct}
{\mcitedefaultendpunct}{\mcitedefaultseppunct}\relax
\EndOfBibitem
\bibitem[Pitarke \latin{et~al.}(2006)Pitarke, Silkin, Chulkov, and
  Echenique]{Pitarke_2006}
Pitarke,~J.~M.; Silkin,~V.~M.; Chulkov,~E.~V.; Echenique,~P.~M. Theory of
  surface plasmons and surface-plasmon polaritons. \emph{Reports on Progress in
  Physics} \textbf{2006}, \emph{70}, 1--87\relax
\mciteBstWouldAddEndPuncttrue
\mciteSetBstMidEndSepPunct{\mcitedefaultmidpunct}
{\mcitedefaultendpunct}{\mcitedefaultseppunct}\relax
\EndOfBibitem
\bibitem[Barnes \latin{et~al.}(2003)Barnes, Dereux, and Ebbesen]{Barnes2003}
Barnes,~W.~L.; Dereux,~A.; Ebbesen,~T.~W. Surface plasmon subwavelength optics.
  \emph{Nature} \textbf{2003}, \emph{424}, 824--830\relax
\mciteBstWouldAddEndPuncttrue
\mciteSetBstMidEndSepPunct{\mcitedefaultmidpunct}
{\mcitedefaultendpunct}{\mcitedefaultseppunct}\relax
\EndOfBibitem
\bibitem[Homola \latin{et~al.}(1999)Homola, Yee, and
  Gauglitz]{HOMOLA1999_SPR_sensors_review}
Homola,~J.; Yee,~S.~S.; Gauglitz,~G. Surface plasmon resonance sensors: review.
  \emph{Sensors and Actuators B: Chemical} \textbf{1999}, \emph{54},
  3--15\relax
\mciteBstWouldAddEndPuncttrue
\mciteSetBstMidEndSepPunct{\mcitedefaultmidpunct}
{\mcitedefaultendpunct}{\mcitedefaultseppunct}\relax
\EndOfBibitem
\bibitem[Homola(2008)]{Homola08_SPP_Sensor_Chem_Bio}
Homola,~J. Surface Plasmon Resonance Sensors for Detection of Chemical and
  Biological Species. \emph{Chemical Reviews} \textbf{2008}, \emph{108},
  462--493, PMID: 18229953\relax
\mciteBstWouldAddEndPuncttrue
\mciteSetBstMidEndSepPunct{\mcitedefaultmidpunct}
{\mcitedefaultendpunct}{\mcitedefaultseppunct}\relax
\EndOfBibitem
\bibitem[Liedberg \latin{et~al.}(1983)Liedberg, Nylander, and
  Lunström]{LIEDBERG1983299}
Liedberg,~B.; Nylander,~C.; Lunström,~I. Surface plasmon resonance for gas
  detection and biosensing. \emph{Sensors and Actuators} \textbf{1983},
  \emph{4}, 299--304\relax
\mciteBstWouldAddEndPuncttrue
\mciteSetBstMidEndSepPunct{\mcitedefaultmidpunct}
{\mcitedefaultendpunct}{\mcitedefaultseppunct}\relax
\EndOfBibitem
\bibitem[Piliarik and Homola(2009)Piliarik, and Homola]{Piliarik:09}
Piliarik,~M.; Homola,~J. Surface plasmon resonance (SPR) sensors: approaching
  their limits? \emph{Opt. Express} \textbf{2009}, \emph{17},
  16505--16517\relax
\mciteBstWouldAddEndPuncttrue
\mciteSetBstMidEndSepPunct{\mcitedefaultmidpunct}
{\mcitedefaultendpunct}{\mcitedefaultseppunct}\relax
\EndOfBibitem
\bibitem[Baburin \latin{et~al.}(2018)Baburin, Kalmykov, Kirtaev, Negrov,
  Moskalev, Ryzhikov, Melentiev, Rodionov, and Balykin]{Baburin:18}
Baburin,~A.~S.; Kalmykov,~A.~S.; Kirtaev,~R.~V.; Negrov,~D.~V.;
  Moskalev,~D.~O.; Ryzhikov,~I.~A.; Melentiev,~P.~N.; Rodionov,~I.~A.;
  Balykin,~V.~I. Toward a theoretically limited SPP propagation length above
  two hundred microns on an ultra-smooth silver surface. \emph{Opt. Mater.
  Express} \textbf{2018}, \emph{8}, 3254--3261\relax
\mciteBstWouldAddEndPuncttrue
\mciteSetBstMidEndSepPunct{\mcitedefaultmidpunct}
{\mcitedefaultendpunct}{\mcitedefaultseppunct}\relax
\EndOfBibitem
\bibitem[Berini(2019)]{Berini:19}
Berini,~P. Highlighting recent progress in long-range surface plasmon
  polaritons: guest editorial. \emph{Adv. Opt. Photon.} \textbf{2019},
  \emph{11}, ED19--ED23\relax
\mciteBstWouldAddEndPuncttrue
\mciteSetBstMidEndSepPunct{\mcitedefaultmidpunct}
{\mcitedefaultendpunct}{\mcitedefaultseppunct}\relax
\EndOfBibitem
\bibitem[Man \latin{et~al.}(2015)Man, Shi, Zhang, Zhang, Min, and
  Yuan]{Man2015}
Man,~Z.; Shi,~W.; Zhang,~Y.; Zhang,~C.; Min,~C.; Yuan,~X.-C. Properties of
  surface plasmon polaritons excited by generalized cylindrical vector beams.
  \emph{Applied Physics B} \textbf{2015}, \emph{119}, 305--311\relax
\mciteBstWouldAddEndPuncttrue
\mciteSetBstMidEndSepPunct{\mcitedefaultmidpunct}
{\mcitedefaultendpunct}{\mcitedefaultseppunct}\relax
\EndOfBibitem
\bibitem[Zhan(2006)]{Zhan06}
Zhan,~Q. Evanescent Bessel beam generation via surface plasmon resonance
  excitation by a radially polarized beam. \emph{Opt. Lett.} \textbf{2006},
  \emph{31}, 1726--1728\relax
\mciteBstWouldAddEndPuncttrue
\mciteSetBstMidEndSepPunct{\mcitedefaultmidpunct}
{\mcitedefaultendpunct}{\mcitedefaultseppunct}\relax
\EndOfBibitem
\bibitem[Salandrino and Christodoulides(2010)Salandrino, and
  Christodoulides]{Salandrino:10}
Salandrino,~A.; Christodoulides,~D.~N. Airy plasmon: a nondiffracting surface
  wave. \emph{Opt. Lett.} \textbf{2010}, \emph{35}, 2082--2084\relax
\mciteBstWouldAddEndPuncttrue
\mciteSetBstMidEndSepPunct{\mcitedefaultmidpunct}
{\mcitedefaultendpunct}{\mcitedefaultseppunct}\relax
\EndOfBibitem
\bibitem[Schepler \latin{et~al.}(2020)Schepler, Yessenov, Zhiyenbayev, and
  Abouraddy]{STSPP_numerical}
Schepler,~K.~L.; Yessenov,~M.; Zhiyenbayev,~Y.; Abouraddy,~A.~F. Space–Time
  Surface Plasmon Polaritons: A New Propagation-Invariant Surface Wave Packet.
  \emph{ACS Photonics} \textbf{2020}, \emph{7}, 2966--2977\relax
\mciteBstWouldAddEndPuncttrue
\mciteSetBstMidEndSepPunct{\mcitedefaultmidpunct}
{\mcitedefaultendpunct}{\mcitedefaultseppunct}\relax
\EndOfBibitem
\bibitem[Kondakci and Abouraddy(2016)Kondakci, and Abouraddy]{Kondakci:16}
Kondakci,~H.~E.; Abouraddy,~A.~F. Diffraction-free pulsed optical beams via
  space-time correlations. \emph{Opt. Express} \textbf{2016}, \emph{24},
  28659--28668\relax
\mciteBstWouldAddEndPuncttrue
\mciteSetBstMidEndSepPunct{\mcitedefaultmidpunct}
{\mcitedefaultendpunct}{\mcitedefaultseppunct}\relax
\EndOfBibitem
\bibitem[Kondakci and Abouraddy(2017)Kondakci, and Abouraddy]{Kondakci2017}
Kondakci,~H.~E.; Abouraddy,~A.~F. Diffraction-free space--time light sheets.
  \emph{Nature Photonics} \textbf{2017}, \emph{11}, 733--740\relax
\mciteBstWouldAddEndPuncttrue
\mciteSetBstMidEndSepPunct{\mcitedefaultmidpunct}
{\mcitedefaultendpunct}{\mcitedefaultseppunct}\relax
\EndOfBibitem
\bibitem[Kondakci and Abouraddy(2018)Kondakci, and Abouraddy]{Kondakci:18}
Kondakci,~H.~E.; Abouraddy,~A.~F. Self-healing of space-time light sheets.
  \emph{Opt. Lett.} \textbf{2018}, \emph{43}, 3830--3833\relax
\mciteBstWouldAddEndPuncttrue
\mciteSetBstMidEndSepPunct{\mcitedefaultmidpunct}
{\mcitedefaultendpunct}{\mcitedefaultseppunct}\relax
\EndOfBibitem
\bibitem[Kondakci and Abouraddy(2019)Kondakci, and Abouraddy]{Kondakci2019}
Kondakci,~H.~E.; Abouraddy,~A.~F. Optical space-time wave packets having
  arbitrary group velocities in free space. \emph{Nature Communications}
  \textbf{2019}, \emph{10}, 929\relax
\mciteBstWouldAddEndPuncttrue
\mciteSetBstMidEndSepPunct{\mcitedefaultmidpunct}
{\mcitedefaultendpunct}{\mcitedefaultseppunct}\relax
\EndOfBibitem
\bibitem[Bhaduri \latin{et~al.}(2018)Bhaduri, Yessenov, and
  Abouraddy]{Bhaduri:18}
Bhaduri,~B.; Yessenov,~M.; Abouraddy,~A.~F. Meters-long propagation of
  diffraction-free space-time light-sheets. \emph{Opt. Express} \textbf{2018},
  \emph{26}, 20111--20121\relax
\mciteBstWouldAddEndPuncttrue
\mciteSetBstMidEndSepPunct{\mcitedefaultmidpunct}
{\mcitedefaultendpunct}{\mcitedefaultseppunct}\relax
\EndOfBibitem
\bibitem[Bhaduri \latin{et~al.}(2019)Bhaduri, Yessenov, Reyes, Pena, Meem,
  Fairchild, Menon, Richardson, and Abouraddy]{Bhaduri19}
Bhaduri,~B.; Yessenov,~M.; Reyes,~D.; Pena,~J.; Meem,~M.; Fairchild,~S.~R.;
  Menon,~R.; Richardson,~M.; Abouraddy,~A.~F. Broadband space-time wave packets
  propagating 70 m. \emph{Opt. Lett.} \textbf{2019}, \emph{44},
  2073--2076\relax
\mciteBstWouldAddEndPuncttrue
\mciteSetBstMidEndSepPunct{\mcitedefaultmidpunct}
{\mcitedefaultendpunct}{\mcitedefaultseppunct}\relax
\EndOfBibitem
\bibitem[Diouf \latin{et~al.}(2021)Diouf, Harling, Yessenov, Hall, Abouraddy,
  and Toussaint]{Diouf21_STvectorbeams}
Diouf,~M.; Harling,~M.; Yessenov,~M.; Hall,~L.~A.; Abouraddy,~A.~F.;
  Toussaint,~K.~C. Space-time vector light sheets. \emph{Opt. Express}
  \textbf{2021}, \emph{29}, 37225--37233\relax
\mciteBstWouldAddEndPuncttrue
\mciteSetBstMidEndSepPunct{\mcitedefaultmidpunct}
{\mcitedefaultendpunct}{\mcitedefaultseppunct}\relax
\EndOfBibitem
\bibitem[Yessenov \latin{et~al.}(2019)Yessenov, Mach, Bhaduri, Mardani,
  Kondakci, Atia, Alonso, and Abouraddy]{Yessenov19_maxgroupdelaySTbeams}
Yessenov,~M.; Mach,~L.; Bhaduri,~B.; Mardani,~D.; Kondakci,~H.~E.; Atia,~G.~K.;
  Alonso,~M.~A.; Abouraddy,~A.~F. What is the maximum differential group delay
  achievable by a space-time wave packet in free space? \emph{Opt. Express}
  \textbf{2019}, \emph{27}, 12443--12457\relax
\mciteBstWouldAddEndPuncttrue
\mciteSetBstMidEndSepPunct{\mcitedefaultmidpunct}
{\mcitedefaultendpunct}{\mcitedefaultseppunct}\relax
\EndOfBibitem
\bibitem[Ichiji \latin{et~al.}(2022)Ichiji, Kikuchi, Yessenov, Schepler,
  Abouraddy, and Kubo]{Ichiji22_STSPParxiv}
Ichiji,~N.; Kikuchi,~H.; Yessenov,~M.; Schepler,~K.~L.; Abouraddy,~A.~F.;
  Kubo,~A. Observation of ultrabroadband striped space-time surface plasmon
  polaritons. 2022; \url{https://arxiv.org/abs/2202.10504}\relax
\mciteBstWouldAddEndPuncttrue
\mciteSetBstMidEndSepPunct{\mcitedefaultmidpunct}
{\mcitedefaultendpunct}{\mcitedefaultseppunct}\relax
\EndOfBibitem
\bibitem[Drezet \latin{et~al.}(2008)Drezet, Hohenau, Koller, Stepanov,
  Ditlbacher, Steinberger, Aussenegg, Leitner, and
  Krenn]{DREZET08_Leakageradiation}
Drezet,~A.; Hohenau,~A.; Koller,~D.; Stepanov,~A.; Ditlbacher,~H.;
  Steinberger,~B.; Aussenegg,~F.; Leitner,~A.; Krenn,~J. Leakage radiation
  microscopy of surface plasmon polaritons. \emph{Materials Science and
  Engineering: B} \textbf{2008}, \emph{149}, 220--229, E-MRS 2007 Spring
  Conference Symposium A: Sub-wavelength photonics throughout the spectrum:
  Materials and Techniques\relax
\mciteBstWouldAddEndPuncttrue
\mciteSetBstMidEndSepPunct{\mcitedefaultmidpunct}
{\mcitedefaultendpunct}{\mcitedefaultseppunct}\relax
\EndOfBibitem
\bibitem[Wiecha \latin{et~al.}(2019)Wiecha, Al-Daffaie, Bogdanov, Thomson,
  Yilmazoglu, Küppers, Soltani, and Roskos]{Wiecha19_nearfieldSPP}
Wiecha,~M.~M.; Al-Daffaie,~S.; Bogdanov,~A.; Thomson,~M.~D.; Yilmazoglu,~O.;
  Küppers,~F.; Soltani,~A.; Roskos,~H.~G. Direct Near-Field Observation of
  Surface Plasmon Polaritons on Silver Nanowires. \emph{ACS Omega}
  \textbf{2019}, \emph{4}, 21962--21966, PMID: 31891075\relax
\mciteBstWouldAddEndPuncttrue
\mciteSetBstMidEndSepPunct{\mcitedefaultmidpunct}
{\mcitedefaultendpunct}{\mcitedefaultseppunct}\relax
\EndOfBibitem
\bibitem[Zhan(2009)]{Zhan09_cylindricalVBs}
Zhan,~Q. Cylindrical vector beams: from mathematical concepts to applications.
  \emph{Adv. Opt. Photon.} \textbf{2009}, \emph{1}, 1--57\relax
\mciteBstWouldAddEndPuncttrue
\mciteSetBstMidEndSepPunct{\mcitedefaultmidpunct}
{\mcitedefaultendpunct}{\mcitedefaultseppunct}\relax
\EndOfBibitem
\bibitem[Toussaint \latin{et~al.}(2005)Toussaint, Park, Jureller, and
  Scherer]{Toussaint05}
Toussaint,~K.~C.; Park,~S.; Jureller,~J.~E.; Scherer,~N.~F. Generation of
  optical vector beams with a diffractive optical element interferometer.
  \emph{Opt. Lett.} \textbf{2005}, \emph{30}, 2846--2848\relax
\mciteBstWouldAddEndPuncttrue
\mciteSetBstMidEndSepPunct{\mcitedefaultmidpunct}
{\mcitedefaultendpunct}{\mcitedefaultseppunct}\relax
\EndOfBibitem
\end{mcitethebibliography}




\section*{Supplementary materials}
Supplementary Text\\
Figs. S1 to S3\\
References (1)-(3)


\end{document}